\documentclass[]{Liebert_Author}
\usepackage{amsmath}
\usepackage{xcolor}

\newcommand{\chla}{Chl \textit{a} }
\newcommand{\chlb}{Chl \textit{b} }

\title{Thermodynamic limits on oxygenic photosynthesis around M-dwarf stars: Generalized models and strategies for optimization} 

\author{Samir Chitnavis,$^{1,2}$ Thomas J. Haworth,$^{3}$ Edward Gillen,$^{3}$ \\Conrad W. Mullineaux,$^{1}$ Christopher D. P. Duffy$^{1,2}$\\
{$^{1}$School of Biological and Behavioural Sciences,}\\
{Queen Mary University of London, Mile End, London E1 4NS, UK}\\
{$^{2}$Digital Environment Research Institute,}\\ {Queen Mary University of London, Empire House, Whitechapel E1 1HH, UK}\\
{$^{3}$Astronomy Unit, Queen Mary University of London, Mile End Road, London E1 4NS, UK}\\
{$^\ast$corresponding author; E-mail:  c.duffy@qmul.ac.uk}}

\begin{document} 

\maketitle 

 
\begin{abstract}
We explore the feasibility and potential characteristics of photosynthetic light-harvesting on exo-planets orbiting in the habitable zone of low mass stars ($< 1$ M$_{\odot}$). As stellar temperature, $T_{s}$, decreases, the irradiance maximum red-shifts out of the $400 \textrm{nm} \leq \lambda < 750$ nm range of wavelengths that can be utilized by \emph{oxygenic} photosynthesis on Earth. However, limited irradiance in this region does not preclude oxygenic photosynthesis and Earth's plants, algae and cyanobacteria all possess very efficient \emph{light-harvesting antennae} that facilitate photosynthesis in very low light. Here we construct general models of photosynthetic light-harvesting structures to determine how an oxygenic photosystem would perform in different irradiant spectral fluxes. We illustrate that the process of light-harvesting, capturing energy over a large antenna and concentrating it into a small \emph{reaction centre}, must overcome a fundamental \emph{entropic barrier}. We show that a plant-like antenna cannot be adapted to the light from stars of $T_{s}<3400$ K, as increasing antenna size offers diminishing returns on light-harvesting. This can be overcome if one introduces a slight \emph{enthalpic gradient}, to the antenna. Interestingly, this strategy appears to have been adopted by Earth's oxygenic cyanobacteria, and we conclude that \emph{bacterial} oxygenic photosynthesis is feasible around even the lowest mass M-dwarf stars.               
\end{abstract}

\section{Introduction}

Of the roughly 5000 exo-planets that have now been identified \citep{ExoSurvey}, how many of them are capable of supporting life? Since the laws of physics and chemistry are universal, there is presumably some (as yet unknown) general rule-set that determines the feasibility and likely characteristics of exo-biospheres. While establishing these rules is, to say the least, a \emph{hard} problem, the laws of thermodynamics dictate that any biosphere must be founded on autotrophic species. A biosphere cannot be a closed system and the organisms at the base of the food chain cannot, by definition, feed on other forms of life. Earth's current biosphere, highly-diverse and abundant in specialized multi-cellular organisms, is (largely) founded upon oxygenic photosynthesis. This is the process of using light to oxidize $H_{2}O$ and ultimately produce two molecules: \emph{Adenosine triphosphate} (ATP) and the reduced form of \emph{Nicotinamide adenine dinucleotide phosphate} (NADPH). These are, respectively, the energy source and reducing power needed for a myriad of biochemical reaction, including carbon dioxide fixation via the Calvin-Benson cycle \citep{Cardona2019,Soo_oxygenic_cyano,Kiang_Photosynthesis_Astro}. The earliest photosynthetic organisms, however, were \emph{anoxygenic}, oxidising compounds such as $H_{2}S$, $H_{2}$, $Fe(II)$, etc. \citep{ozaki2019anoxygenic}. These species were (and remain today) prokaryotes, single-cell organisms lacking a nucleus and any other membrane-enclosed organelles. Today they contribute less than $1\%$ of the global primary production, though this can locally rise to $\sim 30$ \% in sulphide-rich ancient lakes \citep{Overmann2013} 

It has been proposed that oxygenation of Earth's atmosphere by early oxygenic organisms was a necessary prerequisite for the evolution of multi-cellular life (though there is considerable debate \citep[e.g][]{wood2020origin,mills2014oxygen,butterfield2009oxygen,cole2020co,Bozdag2021}). If so then oxygenic photosynthesis may be a universal requirement for biospheres that support `complex' life. Regardless, it may present the best chance for detecting an exo-biosphere in the near future \citep{Kiang_Photosynthesis_Astro}. The presence of ozone ($O_{3}$), produced by photo-excitation of biotically-maintained $O_2$, would produce distinct bands in atmospheric transmission spectra \citep{Mendillo2018,olson2018handbook,Olson_Seasonal2018,lyons2014rise,Schwieterman2018}. Moreover, widespread oxygenic communities on Earth are responsible for its 'vegetation red edge' (VRE), the depletion of the $400 < \lambda < 750$ nm region in Earth's integrated surface reflectance spectrum due to preferential absorption by said organisms \citep{Sagan1993VRE,Kaltenegger_VRE_2018,Seager2005VRE,Arnold_VRE2002}. It is worth noting, however, that widespread oxygenic photosynthesis may still be present without giving rise to either bio-signature \citep{Cockell_2009,Kaltenegger_VRE_2018}.

\begin{figure*}[ht]
\centering
\includegraphics[width=\linewidth]{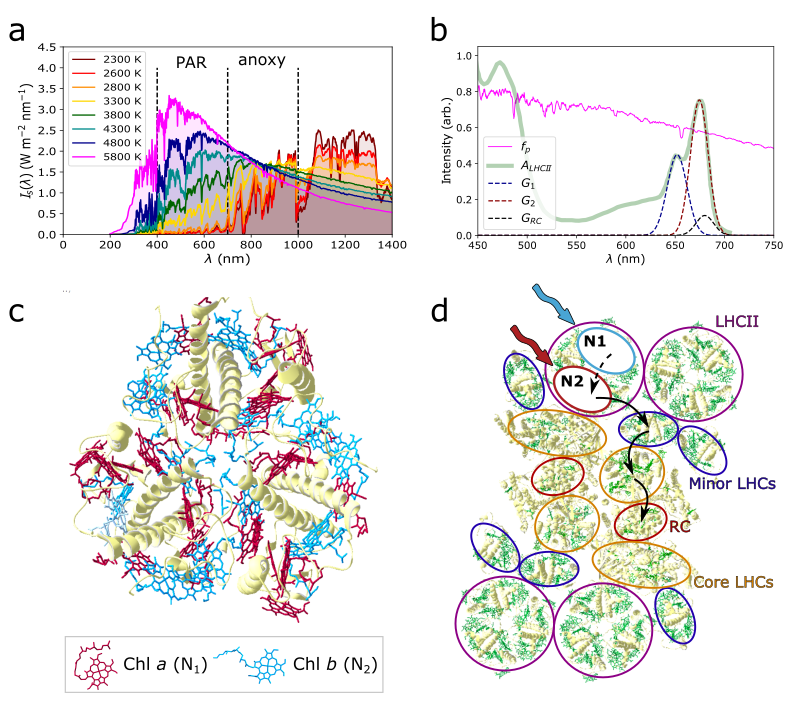}
\caption{\textbf{a.} The incident spectral fluxes, $f_{p}\left(\lambda;T_{s},a_{sp}\right)$ for a range of stellar temperatures, $T_{s}$. The dotted lines indicate the approximate absorption windows for oxygenic (Photosynthetically Active Radiation, PAR) and anoxygenic photosynthesis. \textbf{b.} The absorption profile of major plant antenna complex LHCII (green, digitised from \citep{LHCII_spectrum} using WebPlotDigitizer \citep{Rohatgi2022}) shown with a model Solar ($T_{s}=5800$ K) flux. The red-absorbing $Q_{y}$-band is composed of \chlb ($\sim 650$ nm) and \chla ($\sim 675$ nm) band and are approximated as Gaussian line-shape functions (dotted lines) in subsequent models. The RC absorption maximum (due to redshifted \chla) is at $\sim 680$ nm (black dotted line) and assumed to have a negligible absorption cross-section compared to the antenna. \textbf{c.} An LHCII trimer, viewed from `above' the plane of the membrane in which is sits. \chlb and \textit{a} form two interpenetrating sub-lattices, facilitating rapid ($\sim 1$ ps) energy relaxation from the former to the latter (structure taken from \citep{liu2004crystal}, PDB structure 1RWT). \textbf{d.} The PSII supercomplex as viewed from the same perspective as for the LHCII trimer. A dimer of RC complexes receive energy by a modular assembly of LHCs, including core and minor (monomeric) antenna complexes plus LHCII trimers. Not shown are an additional $\sim 10$ LHCII trimer that are loosely coupled to the supercomplex and form a disordered antenna pool in the membrane. Inter-protein energy transfer (solid arrows) is 1-2 orders of magnitude slower than intra-protein relaxation (\chlb → \chla, dashed arrow).}
\label{Fig1}
\end{figure*}

In addition to all of the basic conditions necessary for life, such as liquid water, bio-available carbon, along with moderate radiation, pressure, etc. \citep{Meadows2018,McKay_Habitable}, oxygenic photosynthesis requires so-called \emph{photosynthetically active radiation} (PAR), photons in the wavelength range $400 <\lambda< 750$ nm (see Fig. \ref{Fig1} \textbf{a}) \citep{McCree1966,McCree1971}. The blue limit reflects the fact that UV-light is photo-damaging to cellular material generally and many photosynthetic organisms produce UV-absorbing compounds such as scytonemin to actively screen against UV over-exposure. The red limit at $700-750$ nm, appears to be enforced by the fundamental redox chemistry of water oxidation, with photons of longer wavelengths being insufficiently energetic to drive the necessary charge-separation processes \citep{Mascoli_FarRed_RC,MascoliRedLimit,Tros2021RedLimit,nurnberg2018photochemistry}. The existence of a hard red limit may then impose strong restrictions on the distribution of oxygenic photosynthesis in our galaxy.  

For some exo-planets it is possible to characterise their masses/radii and hence bulk compositions \cite[e.g.][]{Seager_2007_EoS,Adams_2008_Massive_Atmospheres,Weiss_2014_MassRadius}, and through direct imaging and transmission spectroscopy, probe their atmospheric compositions \cite[e.g.][]{Benneke_2012_Atmos,Nikolov_2016_Atmos, Hinkley_2022_JWST}. Along with the opportunity for detecting bio-signatures, it allows for detailed modelling of the spectral irradiance on the planet surface and therefore the feasibility and possible characteristics of photosynthesis. Stars range in mass from less than $0.1$ to greater than $100$ Solar masses \citep{Wu_2016_stellar_mass}, but follow the stellar \emph{initial mass function} meaning low mass stars are much more common \citep{Kroupa_2001,Chabrier_2003}. Moreover, there is evidence that planets occur more frequently around stars of lower mass than the Sun \citep{Mulders_2015,Hsu_2019_Occurrence,Hsu_2020_Occurrence}, with the planets of Trappist-1 \citep{Gillon_2017_TrappistI},  Proxima Centauri \citep{Anglada-Escude_2016_Proxima, Faria_2022_Proxima} and LHS 1140 \citep{Dittmann_2017_LHS1140} being notable examples. A fraction of these are terrestrial planets orbiting at distances close enough for liquid water to exist on their surfaces \citep{Grimm_2018_Trappist,Dittmann_2017_LHS1140}. However, particularly at the lower end of the stellar mass/temperature range ($T_{s}<3000$ K), the emission peak wavelength of the parent star is red-shifted well below the PAR range (see Fig. \ref{Fig1} \textbf{a}). It may be that the spectral irradiance on the surface of exoplanets orbiting very low mass stars like Trappist-1 is simply insufficient to ``power'' oxygenic photosynthesis, a hypothesis supported by the thermodynamic modelling of Covone et al.\citep{CovoneOxygenic2021}. 

On Earth, there is considerable diversity in photosynthesis, reflecting huge variability in local light intensity and spectral quality. Vascular plants, mosses and green algae use chlorophyll \textit{a} and \textit{b} (Chl \textit{a} and \textit{b}) which absorb strongly in the red ($650-700$ nm) and blue ($350-500$ nm) regions of the Solar spectrum (see Fig. \ref{Fig1} \textbf{b}) \citep{KnoxSpring2003}. Cyanobacteria and red algae, which are also oxygenic, use \chla but also bind pigments such as phycocyanobilin ($620-650$ nm) and phycoerythrobilin ($560-580$ nm), giving them much broader spectral coverage than plants. Some species of deep-water marine cyanobacteria also bind phycourobilin ($\sim 495$ nm), an adaptation of the predominantly blue light that penetrates to those depths \citep{saer2017light,Kolodny_2022_Phyco_blue}. Yet, other cyanobacteria bind \emph{red-shifted} Chl \textit{d} and \textit{f} ($750-800$ nm), possibly to take advantage of the scarce red light not prioritized by their cousins \citep{ViolaRedPhoto2022,Tros2021RedLimit}. There are also numerous anoxygenic prokaryotes that utilize redder photons in the $\lambda\sim 800-900$ nm and even $\lambda\sim 1000$ nm ranges \citep{Bryant2006Prokaryote}. They bind various Bacteriochlorophylls (BChls) and include groups like the purple bacteria \citep{hu_ritz_damjanović_autenrieth_schulten_2002} and green sulphur bacteria \citep{Gregersen_Green_Sulfur}. 

However, despite this considerable diversity, all photosynthetic organisms utilize an evolutionary strategy known as the \emph{antenna-reaction centre architecture} \citep{WolfeAntenna1994,FlemingPrinciples2012}. This is a division of labour in which a small subset of pigments are incorporated into the \emph{reaction centres} (RCs), the specialized proteins that carry out the first steps of the photosynthetic light reactions (photo-oxidation of a substrate, reduction of a mobile electron carrier). The remaining pigments (the vast majority) are bound to large modular assemblies of light-harvesting or \emph{antenna} complexes (LHCs, see Fig. \ref{Fig1} \textbf{c}) which capture light and transfer the resulting excitations to the RCs (Fig. \ref{Fig1} \textbf{d}). LHCs generally bind several types of pigments in order to provide broad spectral coverage and create a `funnel' structure in which higher energy pigments donate to lower energy ones. The antenna evolved to facilitate photosynthesis in very low light, an extreme example being some species of green sulphur bacteria which utilize the faint black-body radiation emitted by deep-sea hydrothermal vents \citep{BeattyGSBhydrothermal}. Nevertheless LHCs are not black bodies but rather absorb over selective (and often quite narrow) bands.

Elucidating a general set of rules that quantitatively relates the absorption properties of antennae to the local light environment in which they evolved is an active topic. Kiang et al. derived a set of empirical rules based on general observation \citep{Kiang2007}: (1) The absorption peak of the antenna is close to the local irradiance maximum. (2) The absorption peak of the RCs is close to the longest wavelength in the irradiance range. (3) `Secondary' pigments (such as Chl \textit{b} or carotenoids in plants) will absorb towards the shortest wavelength of the irradiance window. \cite{Bjorn1976} and later \citet{Marosvolgyi2010} considered the balance between the need to absorb as much light as possible with the potentially prohibitive metabolic cost of synthesizing and maintaining a vast array of different pigment co-factors. This neatly predicts the very different absorption maxima of both plants and purple bacteria. More recently, \cite{Arp2020} argued that the absorption profiles of LHCs are optimized to be robust against fluctuations in both irradiance (termed ``external noise'') and pigment-to-pigment energy transfer (``internal noise''). In effect this requires the that LHCs bind two similar pigments (e.g. Chl \textit{a} and \textit{b}) with close absorption maxima centred on the steepest part of the local spectral irradiance (Fig. \ref{Fig1} \textbf{b}). 

These various rule-sets have been applied to spectral irradiances from lower mass stars. Wolstencroft and Raven simply took the action spectrum of the plant \emph{Nerium oleander} and applied it to a range of model spectral irradiances, concluding oxygenic photosynthesis performs best around F-type stars ($T_{s}=6000-7500$ K) and very poorly around around K and M-type ($T_{s}<5200$ K)\citep{Wolstencroft2002}. Similarly, \citet{CHall_Habitable_Preprint} took an action spectrum measured for phytoplankton \citep{Yang2020Phytoplankton}) and concluded that, given otherwise favourable conditions, oxygenic photosynthesis may be possible for larger K-type stars. Kiang et al applied their empirical rule-set to Earth-like planets orbiting F, K and M-type stars and predicted that the latter would favour LHCs that absorb in the NIR ($930-2500$ nm) which have no direct analogue on Earth \citep{Kiang2007_Part2}. \citet{LehmerFarRedEdge2021} applied Marosv\"{o}lgyi's and van Gorkom's model to the same stellar spectra, predicting a progressive red-shift in the optimal antenna with decreasing $T_{s}$, arriving at the NIR ($\sim 1000$ nm) for the lowest mass M-dwarfs. A similar trend was predicted by \citet{duffy2023photosynthesis} who applied a modified version of the noise cancelling criteria of \citet{Arp2020}, suggesting that anoxygenic photosynthesis would be favourable around M-dwarf stars. Lastly, \citet{Lingam2021} correlated $T_{s}$ to the $pi$-electron conjugation length that a pigment molecule would need to most effectively absorb light, assuming the pigment has an optical gap tuned to the stellar emission peak. They concluded that these pigments would need roughly twice the number of $\pi$-electrons for $T_{s}<3000$ K compared to $T_{s}\sim 6000$ K, again supporting the idea that the emission spectra of low mass stars may select for NIR-absorbing, anoxygenic organisms. 

All of these models focus on the intrinsic optical properties of pigment co-factors or assume a fixed action spectrum, making no assumptions regarding the antenna superstructure. This neglects two defining characteristics of antenna systems: That they initially evolved to maximize light capture in low light environments and, through their modular structure, they are highly adaptable to changes in light \citep{Lokstein2021_review}. Hereafter we will focus exclusively on Photosystem II (PSII) which is the water-oxidizing photosystem in vascular plants, algae, mosses, and cyanobacteria. It is PSII that maintains the $O_{2}$ content of Earth's atmosphere and therefore produces an obvious bio-signature. While the PSII is more-or-less conserved across all oxygenic species, the light-harvesting antenna varies considerably. We will, initially atleast, focus on the PSII antenna of vascular plants, since this accounts for roughly half the chlorophyll on Earth \citep{jansson1994light} and is the foundations of the Earth's forests, savannas and arable land. The basic unit of plant PSII antenna is LHCII, a cyclic trimer shown in Fig. \ref{Fig1} \textbf{a}. It binds both Chl \textit{a} and Chl \textit{b}, with the latter transferring energy to the former, plus several green-absorbing carotenoids \citep{liu2004crystal}. LHCII, plus some monomeric (``minor'') variants, bind to a dimer of PSII RC complexes to form the \emph{PSII supercomplex} which is shown in Fig. \ref{Fig1} \textbf{d} \citep{KOURIL20122,Liu_LHCII_PSII}. Not shown is the pool of additional, loosely-bound LHCII trimers that make up the disordered peripheral PSII antenna. The process of light-harvesting and conversion proceeds (approximately) as follows:-
\begin{enumerate}
\item{A \chlb or other accessory pigment absorbs a photon and the resulting excitation rapidly ($\sim 1$ ps) relaxes to Chl \textit{a}. Direct excitation of \chla is also likely \citep{novoderezhkin_intra-_2011}.}
\item{The excitation equilibrates ($\sim 1$ ps) across the \chla molecules of the LHCII trimer \citep{novoderezhkin_intra-_2011}.}
\item{Over $\sim10-100$ ps the excitation ``hops'' stochastically between different antenna complexes, following a (very slight) energetic gradient towards the RC \citep{Valkunas_CG_2009}.}
\item{The excitation arrives on the PSII Chl \textit{a} \emph{special pair} (given the spectroscopic name P680 due to its fluorescence maximum) and is converted into a meta-stable intermolecular \emph{charge-separated state}. This is known as \emph{primary trapping} and takes $\sim 5$ ps \citep{Broess_2006}.}
\item{The charge separated state reduces a mobile electron carrier (a quinone) which initiates the photosynthetic electron transport chain. This is the rate limiting step, typically taking $\sim 10$ ms \citep{oja2000oxygen}.}
\end{enumerate}
This is a highly-efficient process, with roughly 85\% of the photons captured by the antenna going on the generate an electron in the RC \citep{Baker_2004}. Moreover, the PSII antenna is extremely adaptable, being able to alter its size, efficiency and connectivity to the RC in response to changing light levels. For sustained periods of limited illumination the organism will ``acclimate'' by (among a range of physiological responses) expressing more LHCII \citep{Bailey_2001} and modulating protein packing to increase the coupling between PSII and the peripheral antenna \citep{Kouril2013}. This is not unique to plants and while the antennae of, say, cyanobacteria differ strongly in terms of structure, they operate on the same basic principles.

With the concept of the antenna introduced it seems reasonable that hypothetical oxygenic organisms could compensate for limited PAR emitted by an ultra low mass star by simply evolving a very large antenna, composed of lots of LHCs. However, a larger antenna comes at a cost, most obviously in increased metabolic burden of synthesizing large amounts of pigments and protein. Setting this aside, there is potentially a more fundamental limit. Light-harvesting essentially involves generating excitations (i.e. capturing photons) over a large area and then concentrating them into a small one, a process which \emph{decreases} local entropy. In certain cases (see Methods) this will impose a thermodynamic barrier to light-harvesting that will only get steeper for a larger antenna. In this work, using a series of generalized antenna models, we explore how a fundamental thermodynamic barrier may restrict the effectiveness of plant-like oxygenic photosynthesis around low mass stars. We then discuss general evolutionary strategies for over-coming the barrier and how these have already been realized in Earth's oxygenic cyanobacteria.    

\section{Methodology}

\subsection{Local spectral irradiance as a function of stellar temperature}

As in previous work \citep{duffy2023photosynthesis}, we use stellar spectral models for stars of different effective temperatures, $T_{s}$, generated by the \textsc{phoenix} code \citep{Husser_PHOENIX_2013}. To reduce expense of subsequent numerical integration calculations, we smooth and re-sample the spectrum down to 4000 points, which still captures the large scale features. The absorption profiles of typical light-harvesting pigments are quite broad, with a \emph{Full Width at Half Maximum} (FWHM) $> 10$ nm. Photon absorption by these pigments therefore integrates over this wavelength range (see next subsection), meaning that the photosynthetic pigments can only \emph{resolve} spectral details on scales comparable to or larger than their own absorption line-width. Even with this smoothing, spectral resolution is well below this limit. 

We relate the stellar surface spectral flux density, $I_{s}\left(\lambda;R_{s},T_{s}\right)$ to the spectral flux density at the planetary surface, $f_{p}\left(\lambda ;R_{s},T_{s},a_{sp}\right)$, with the radiative equilibrium condition,
\begin{equation}\label{eq:1}
f_{p}\left(\lambda;R_{s},T_{s},a_{sp}\right) =
\left(\frac{R_{s}}{a_{sp}}\right)^{2} I_{s}\left(\lambda;R_{s},T_{s}\right)
\end{equation}
where $R_{s}$ is the stellar radius, $a_{sp}$ is the mean orbital distance of the planet and both $I_{s}\left(\lambda\right)$ and $f_{p}\left(\lambda\right)$ have units of W m$^{-2}$ nm$^{-1}$. During the stellar main sequence $R_{s}$ and $T_{s}$ are related by a power law and 
and therefore we hereafter specify $I_{s}(\lambda;T_{s})$ and $f_{p}\left(\lambda;T_{s},a_{sp}\right)$ in terms of $T_{s}$ only. 
We consider $2300 K \leq T_{s} \leq 5800 K$ , with $T_{s} = 2300 K$ representing the coolest end of the M-dwarf stellar range and $T_{s} = 5800 K$ giving an approximation of the spectral flux density of the Sun. We obtain approximate habitable orbital distances using the relation, 
\begin{equation}\label{eq:2}
a_{sp}=\left(\frac{R_{s}^{2}}{4}\left(\frac{T_{s}}{T_{p}}\right)^{4}\right)^{\frac{1}{2}}
\end{equation}
where $T_{p}$ is the average planetary surface temperature, taking the mid-point between the values of $a_{sp}$ that yield $T_{p}=273$ K and $373$ K.
Note, that for reasons of generality we do not consider any albedo (reflectance) or atmospheric attenuation of $f_{p}\left(\lambda;T_{s},a_{sp}\right)$. The possible implications of this are addressed in the discussion.  

\subsection{Thermodynamic model of an oxygenic light-harvesting system}

\begin{figure*}[ht]
\centering
\includegraphics[width=\linewidth]{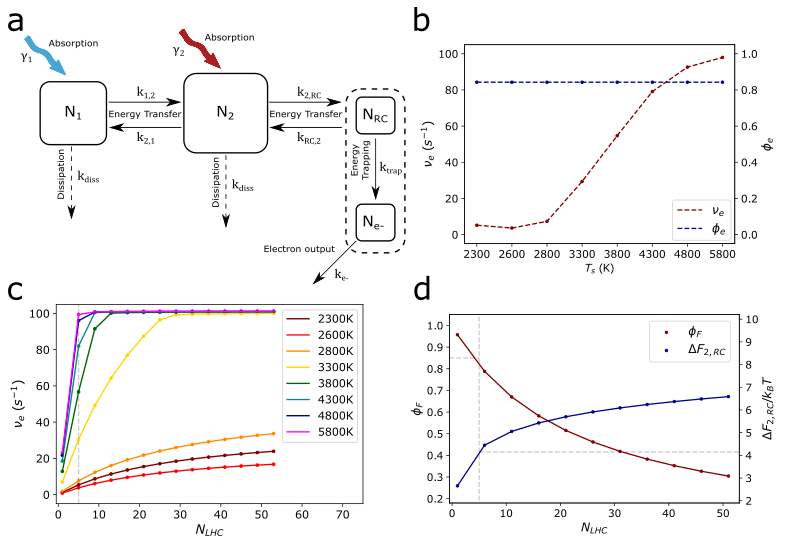}
\caption{\textbf{a.} Schematic of the thermodynamic antenna model.  $N_{1}$ and $N_{2}$  represent the two absorber populations (\chlb and \textit{a} respectively in plants) which capture photos with rate constants $\gamma_{1}$ and $\gamma_{2}$ respectively. The $N_{1}\rightleftarrows N_{2}$ transfer step represents fast ($\sim 1$ ps) equilibration between these domains within a single LHC. $N_{2}\rightleftarrows N_{RC}$ represents longer range transfer to the RC. The final step represents irreversible ‘trapping’ of the exciton in which it is rapidly ($\sim5$ ps) converted into a charge-separated state. ‘Electron output’ denotes the reduction of the electron acceptor (a quinone for PSII). \text{b.} The electron production rate, $\nu_{e}$, of the same model under different spectral fluxes,$T_{s}$. The model is tuned to give the maximal PSII rate ($\nu_{e}\sim 100$ electron $s^{-1}$) for $T_{s}=5800$ K. Also shown is the RC quantum efficiency, $\phi_{e}$, which is independent of flux and matches the $\phi_{e}=0.8 - 0.85$ observed for PSII in plants. \textbf{c.} $\nu_{e}$ as a function of increasing antenna size, $N_{LHC}$, and $T_{s}$. \textbf{d.} The free energy increase, $\Delta F_{2,RC}$, associated with transfer from the larger antenna pool to the RC (blue) and $\phi_{e}$ as a function of $N_{LHC}$. Both quantities are independent of $\gamma_{1,2}$ and therefore independent of $T_{s}$. The dashed line denote the values for the $N_{LHC}=5$ model.}
\label{Fig2}
\end{figure*}

Our first light-harvesting model is general, explicitly thermodynamic and based \emph{loosely} on PSII of vascular plants. However, the exercise is not to precisely simulate the light-harvesting dynamics of plants but to compare the performance of a qualitative model in different spectral irradiances. A schematic of the model is shown in Fig. \ref{Fig2} \textbf{a}. We assume that the antenna is composed of two energetically distinct \textbf{domains} which in plants are the interpenetrating \chla and \textit{b} pools. At time, $t$, these two pools contain $N_{1}\left(t\right)$ and $N_{2}\left(t\right)$ excitations respectively. Energy is transferred sequentially from $N_{1}$ to $N_{2}$ to the RC, whose excitation occupation is denoted $N_{RC}\left(t\right)$. The RC then \textit{traps} this excitation by converting it into a charge separated state, which we assume is irreversible. The \textit{trap states}, $N_{e^{-}}\left(t\right)$, represents the  meta-stable charge-separated state that eventually reduces the (quinone) electron acceptor. The equations of motion for this system are,
\begin{equation}\label{eq:3}
\frac{d}{dt}N_{1}=-\left(k_{1,2}+k_{diss}\right)N_{1}+k_{2,1}N_{2}+\gamma_{1}
\end{equation}
\begin{equation}\label{eq:4}
\frac{d}{dt}N_{2}=k_{1,2}N_{1}-\left(k_{2,1}+k_{2,RC}+k_{diss}\right)N_{2}+k_{RC,2}N_{RC}+\gamma_{2}
\end{equation}
\begin{equation}\label{eq:5}
\frac{d}{dt}N_{RC}=k_{2,RC}N_{2}-\left(k_{RC,2}+k_{trap}\right)N_{RC}
\end{equation}
\begin{equation}\label{eq:6}
\frac{d}{dt}N_{e}=k_{trap}N_{RC}-k_{e}N_{e}
\end{equation}
where $k_{i,j}$ denotes the rate constant for excitation transfer \textit{from} domain $i$ \textit{to} domain $j$, $k_{x}$ denotes a rate of excitation loss due to some decay/conversion process, and $\gamma_{i}$ denotes the rate of photon absorption by domain $i$. 

We first consider empirical decay/conversion rate constants. $k_{diss}$ is the intrinsic excited state dissipation rate of the pigments in the antenna, which for \chla and \textit{b} bound to photosynthetic proteins is $k_{diss}^{-1}\sim 4$ ns \citep{Barzda_2001}. $k_{trap}$ is the rate of primary charge separation, $k_{trap}^{-1}\sim 5$ ps \citep{Broess_2006}, which we assume is irreversible \citep{Valkunas_CG_2009}. $k_{e}$ is the rate of reduction of the electron acceptor. In PSII this is a 2-electron process in which plastoquinone is reduced to plastoquinol,
\begin{equation}\label{eq:7}
PQ + 2H^{+} + 2e^{-}\xlongrightarrow{\gamma} PQH_{2}
\end{equation}
but here set $k_{e^{-}}^{-1} = 10$ ms as this reflects the typical maximal working rate of PSII (roughly $100$ photons converted into $100$ electrons per second \citep{oja2000oxygen}).

Next we define the photon input rates $\gamma_{1}$ and $\gamma_{2}$. With reference to the plant PSII (Fig. \ref{Fig1} \textbf{d}) we assume the RC is coupled to a pool of $N_{LHC}$ identical LHC that each contain $n_{1}$ \chlb and $n_{2}$ \chla pigments. We assume that all pigments of a given type are \emph{thermodynamically equivalent}, meaning that the excitation can sample them all equally and instantaneously. In effect this requires them to have the same absorption peak wavelength, absorption width and excited state lifetime. A single LHCII trimer contains (on average) $n_{1}=18$ \chlb and $n_{2}=24$ \chla \citep{liu2004crystal}. In reality, the pigments in each pool are not strictly equivalent as local interactions with their respective protein binding pockets induce small shifts in excitation energy. One can define the \emph{effective} number of equivalent states,
\begin{equation}\label{eq:8}
\tilde{n_{i}}=\sum_{m\in i}\textrm{exp}\left(-\frac{hc}{\lambda_{m}^{p}k_{B}T}\right)
\end{equation}
where $\lambda_{m}^{p}$ is the peak wavelength of pigment $m$ (which belongs to domain $i$), $k_{B}$ is Boltzmann's constant, and $T$ is local temperature. For simplicity we will assume $n_{1}=18$, $n_{2}=24$ and $T=300$ K throughout. In plant PSII $N_{LHC}$ is variable depending on species and light-environment but there are roughly $5$ LHCII trimers per RC core, plus three minor (monomeric) LHCII complexes and two core antennae \citep{DEKKER200512}. We treat $N_{LHC}$ as a variable parameter throughout. 

We assume domains $N_{1}$ and $N_{2}$ have Gaussian absorption profiles,
\begin{equation}\label{eq:9}
G_{i}\left(\lambda\right)=n_{i}N_{LHC}\sigma_{i}\frac{1}{w_{i}\sqrt{2\pi}}\textrm{exp}\left(-\frac{\left(\lambda-\lambda_{i}^{p}\right)^{2}}{2w_{i}^{2}}\right)
\end{equation}
where $w_{i}$ is the standard deviation (hereafter termed the ``width''), and $\sigma_{i}$ is an absorption cross-section. We assume $\sigma_{1}=B_{1,2}\tilde{\sigma}$ and $\sigma_{2} = \tilde{\sigma}$ where $B_{1,2}$ encodes any \emph{intrinsic} differences in the optical properties of the two pigment domains. The parameters $(B_{1,2},\lambda_{p}^{1}, w_{1}, \lambda_{2}, w_{2})$ are obtained by visual fitting of the Chl \textit{a} and \textit{b} $Q_{y}$-peaks in the LHCII absorption profile (Fig. \ref{Fig1} \textbf{b}). We treat $\tilde{\sigma}$, as a free parameter. The photon absorption rates are therefore,
\begin{equation}\label{eq:10}
\gamma_{i}=\int_{0}^{\infty}\frac{\lambda}{hc}f_{p}\left(\lambda;T_{s},a_{sp}\right)G_{i}\left(\lambda\right)d\lambda 
\end{equation}
It is assumed that the RC has a the same line-shape as $N_{2}$ (since it is also composed of Chl \textit{a}) but red-shifted to $\lambda_{RC}^{p}=680$ nm and with $\sigma_{RC}\sim 0$ \citep{Cardona2012}. 

Lastly, we can define the $N_{1}\rightleftarrows N_{2}$ and $N_{2}\rightleftarrows N_{RC}$ transfer rates and it is here where thermodynamics explicitly enter the model. In addition to the equivalence of pigment states within each domain we also assume that:-
\begin{enumerate}
\item{Excitation occupation is low at any given point, $N_{i}(t)<<n_{i}N_{LHC}$, meaning we neglect multi-excitation effects such as annihilation or entropic repulsion.}
\item{Equilibration of the excitation within a domain is much faster than inter-domain transfer.}
\end{enumerate}
The transfer rates therefore satisfy the detailed balance condition,  
\begin{equation}\label{eq:11}
\frac{k_{i,j}}{k_{j,i}}=\textrm{exp}\left(-\frac{\Delta F_{i,j}}{k_{B}T}\right)
\end{equation}
where $\Delta F_{i,j}$ is the (Helmholtz) free energy change associated with the transfer of an excitation \emph{from} $N_{i}$ \emph{to} $N_{j}$ transfer. $\Delta F_{i,j}$ is defined, 
\begin{equation}\label{eq:12}
\Delta F_{i,j}=H_{i,j}-T\Delta S_{i,j}
\end{equation}
where the enthalpy change is determined by the energy difference between domains,
\begin{equation}\label{eq:13}
\Delta H_{i,j}=hc\left(\frac{1}{\lambda_{i}^{p}}-\frac{1}{\lambda_{j}^{p}}\right)
\end{equation}
and the entropy change is determined by the difference in the number of equivalent states in the two domains, 
\begin{equation}\label{eq:14}
\Delta S_{i,j}=k_{B}\ln{\left(\frac{\tilde{n}_{j}}{\tilde{n}_{i}}\right)}
\end{equation}
If $\Delta F_{i,j}<0$ then,
\begin{equation}\label{eq:15}
\left(k_{i,j},k_{j,i}\right)=\left(K_{ij}, K_{ij}\textrm{exp}\left[-\frac{\Delta F_{j,i}}{k_{B}T}\right]\right)
\end{equation}
where $K_{ij}$ is some \emph{intrinsic} transfer rate, and $\Delta F_{j,i}=-\Delta F_{i,j}$. In Eqn. (\ref{eq:15}) the $N_{i}\rightarrow N_{j}$ step is thermodynamically favourable and so is only limited by intrinsic parameters such as the resonant coupling between pigments, their spectral overlap, etc. \citep{novoderezhkin_intra-_2011}, all of which are absorbed into $k_{i,j}=K_{ij}$.  The \emph{reverse} transfer ($N_{j}\rightarrow N_{i}$) is unfavourable and therefore subject to an additional \emph{Boltzmann penalty} which is encoded in the factor of $\textrm{exp}\left(-\Delta F_{j,i}/k_{B}T\right)<<1$. 

For plant PSII $N_{1}\rightleftharpoons N_{2}$ corresponds to the intra-LHCII \chlb $\rightarrow$ \chla relaxation step and so $K_{1,2}\sim 1$ ps \citep{novoderezhkin_intra-_2011} (see Fig. \ref{Fig1} \textbf{d}). $N_{2}\rightleftharpoons N_{RC}$ involves long-range, inter-LHCII hopping (\ref{Fig1} \textbf{d}) which is slower and involves more pigments than the previous step. While it is enthalpically favourable (but less than $N_{1}\rightarrow N_{2}$) it is very entropically \emph{unfavourable}. If we assume that the RC contains a single pigment state (since it can ``process'' only one photon at a time \citep{Cardona2019}) then,
\begin{equation}\label{eq:16}
   k_{B} \ln\left(\frac{1}{n_{2}N_{LHC}}\right) < \Delta S_{2,RC} <0
\end{equation}
where the lower limit assumes that the excitation equilibrates across the whole antenna before transfer to the RC. In reality, particularly for large $N_{LHC}$, the excitation cannot equally sample two distantly separated LHCs, particularly if they are on opposites sides of the RC as in the PSII supercomplex (Fig. \ref{Fig1} \textbf{d}). For model simplicity we assume the lower limit throughout and interpret our results accordingly.

If we assume the steady state ($dN_{i}/dt=0$ for all $i$) the problem becomes solving, 
\begin{equation}\label{eq:17}
\begin{pmatrix}
-\left(k_{1,2}+k_diss\right) & k_{21} & 0 & 0\\
k_{1,2} & -\left(k_{2,1}+k_{2,RC+k_{diss}}\right) & k_{RC,2} & 0 \\
0 &  k_{2,RC} & -\left(k_{RC,2}+k_{trap}\right) & 0\\
0 & 0 & k_{trap} & -k_{e}
\end{pmatrix}
\begin{pmatrix}
N_{1}^{eq}\\ N_{2}^{eq}\\N_{RC}^{eq}\\ N_{e}^{eq}
\end{pmatrix}
=
\begin{pmatrix}
-\gamma_{1}\\ -\gamma_{2}\\ 0\\0
\end{pmatrix}
\end{equation}
for the equilibrium excitation populations $\left(N_{1}^{eq},N_{2}^{eq},N_{RC}^{eq},N_{e}^{eq}\right)$. We can then derive two \emph{observable} quantities: The electron output rate of the RC,
\begin{equation}\label{eq:18}
    \nu_{e}=k_{e}N_{e}^{eq}
\end{equation}
which can be measured (indirectly) using, for example, an oxygen electrode \citep{oja2000oxygen}, and the photon-to-electron quantum efficiency,
\begin{equation}\label{eq:19}
    \phi_{e}=\frac{\nu_{e}}{\nu_{e}+k_{diss}\left(N_{1}^{eq}+N_{2}^{eq}\right)}
\end{equation}
which is measured for PSII using techniques such as Pulse Amplitude Modulation (PAM) Fluorescence \citep{Semer_PAM_2019}. For plant PSII the maximal/optimal output rate is $\nu_{e}\sim 100$ s$^{-1}$ \citep{oja2000oxygen} while $\phi_{e} \sim 0.80-0.85$ \citep{Baker_2004}. 

Finally, taking all of these parameters (which are listed in Table \ref{tab:1}) we adjust $\tilde{\sigma}$ and $N_{LHC}$ until the model yields $\nu_{e}\sim 100$ s$^{1}$ and $\phi \sim 0.85$ when subject to $f_{p}\left(\lambda;T_{s}=5800 \textrm{K}\right)$. Keeping this set of parameters we vary $T_{s}$ then attempt to recover Earth-like photosynthetic performance \textbf{by increasing $N_{LHC}$ only}. 

\begin{table}[!h]
\begin{center}
\caption{\label{tab:1} Fixed parameters for the thermodynamic model of a generalized oxygenic light-haveresting system.}
\begin{tabular}{ |c|c|c|c| }
 \hline
 Parameter & Denotes & Value & Ref. \\
 \hline
 \hline
 $k_{diss}^{-1}$ & Pigment excitation lifetime & $4$ ns & \cite{Barzda_2001} \\ 
 $k_{trap}^{-1}$ & RC trapping time & $5$ ps & \cite{Broess_2006} \\ 
 $k_{e^{-}}^{-1}$ & RC electron turnover time & $10$ ms & \cite{oja2000oxygen}\\
 \hline
 $(n_{1},n_{2},n_{RC})$ & Number of pigments per subunit & $(18,24,1)$ & \cite{Liu_LHCII_PSII}\\
 $(B_{1,2},\lambda_{1}^{p},w_{1})$ & $N_{1}$ absorption parameters & $(0.56,650 \textrm{ nm}, 8.5 \textrm{ nm})$ & \cite{LHCII_spectrum,Rohatgi2022}\\
 $(\lambda_{2}^{p},w_{2})$ & $N_{2}$ absorption parameters & $(675 \textrm{ nm},9.0 \textrm{ nm})$ & \cite{LHCII_spectrum,Rohatgi2022}\\
 $(\sigma_{RC},\lambda_{RC}^{p},w_{RC})$ & RC line-shape parameters & $(0,680 \textrm{ nm},9.0 \textrm{ nm})$ & \cite{Cardona2012}\\
 \hline
 $K_{12}^{-1}$ & Intrinsic $N_{1}\rightleftharpoons N_{2}$ transfer time & $1$ ps & \cite{novoderezhkin_intra-_2011}\\
 $K_{2RC}$ & Intrinsic $N_{2}\rightleftharpoons N_{RC}$ transfer time & $10$ ps & \cite{Valkunas_CG_2009} \\
 \hline
\end{tabular}
\end{center} 
\end{table}

\subsection{Lattice model of an oxygenic light-harvesting system}

While the previous model explicitly relates $\nu_{e}$ and $\phi_{e}$ to the thermodynamic properties of the antenna, it neglects the influence of antenna topology. The assumption that the excitation samples the entirety of the antenna super-structure is unrealistic, as is the assumption of a single rate constant for $N_{2}\rightarrow N_{RC}$ (see Fig. \ref{Fig1} \textbf{d}). Therefore, we construct a complimentary lattice diffusion model in which localized, non-interacting, excitations ``hop'' between clusters of pigments. Specific antenna topologies are discussed in the Results section but in essence a set of discrete antenna ``sites'' are arranged around a central RC, either in linear branches (see Fig. \ref{Fig3} \textbf{a}) or a hexagonal lattice (see Fig. \ref{Fig3} \textbf{b}). The sites can be pigment clusters within LHCs, entire LHCs, or even blocks of multiple LHCs, but the assumption is always that excitation equilibration within a site is much faster than inter-site hopping. For $N_{s}$ antenna sites we define the population vector as, 
\begin{equation}\label{eq:20}
\vec{\textbf{N}}\left(t\right)=\left(N_{0}(t),N_{1}(t),\cdots,N_{m}(t),\cdots,N_{N_{s}+1}(t)\right)^{\intercal}
\end{equation}
where $N_{0}$ the charge-separated trap state and $N_{1}$ is the RC. The topology of the antenna is defined by an adjacency matrix, $\textbf{A}$, where $A_{mn}=1$ if sites $m$ and $n$ can exchange energy and $A_{mn}=0$ otherwise. The equations of motion are conveniently expressed in matrix form,
\begin{equation}\label{eq:21}
\frac{d}{dt}\vec{\textbf{N}}(t)=\boldsymbol{T}\vec{\textbf{N}}(t)+\vec{\boldsymbol{\gamma}}
\end{equation}
where,
\begin{equation}\label{eq:22}
\vec{\boldsymbol{\gamma}}=\left(0,0,\gamma_{1},\cdots,\gamma_{m},\cdots,\gamma_{N_{s}}\right)^{\intercal}
\end{equation}
is the vector of photon input rates for each site and $\boldsymbol{T}$ is the \emph{transfer matrix}. The off-diagonal elements of $\textbf{T}$ are defined,
\begin{equation}\label{eq:23}
T_{mn}=
\begin{cases}
k_{trap}, & \textrm{for } m=0 \textrm{, } n=1\\
0, &\textrm{for } m=1 \textrm{, } n=0\\
k_{n,m}A_{nm}, & \textrm{otherwise}
\end{cases}
\end{equation}
where the second case arises from the fact that trapping is irreversible. The diagonal elements are,
\begin{equation}\label{eq:24}
T_{mm}=
\begin{cases}
-\sum_{n\neq m}T_{nm}-k_{e}, & \textrm{for } m=0\\
-\sum_{n\neq m}T_{nm}-k_{diss}-k_{trap}, &\textrm{for } m=1\\
-\sum_{n\neq m}T_{nm}-k_{diss}, & \textrm{otherwise}
\end{cases}
\end{equation}
All rates are defined as they were in the thermodynamic model. If we define $\Delta F_{m,n}<0$ as the free energy change for $N_{m}\rightarrow N_{n}$ transfer then,
\begin{equation}\label{eq:25}
\left(k_{m,n},k_{n,m}\right)=\left(K_{mn}^{hop},K_{mn}^{hop}\textrm{exp}\left[-\frac{\Delta F_{n,m}}{k_{B}T}\right]\right)
\end{equation}
where $K_{mn}^{hop}$ is an \emph{intrinsic} inter-site hopping rate for a given pair of sites. Note that
\begin{equation}\label{eq:26}
\Delta F_{m,n}=hc\left(\frac{1}{\lambda_{m}^{p}}-\frac{1}{\lambda_{n}{p}}\right)-k_{B}T\ln\left(\frac{\tilde{n}_{n}}{\tilde{n}_{m}}\right)
\end{equation}
as before but now only encodes the thermodynamics of a single hopping event. The overall thermodynamics of energy capture, migration to the RC and trapping are strictly implicit. 

As with the thermodynamic model the dynamics are solved in the steady state to obtain $\nu_{e}$ and $\phi_{e}$. We start with a model that aims to replicate plant PSII for $f_{p}\left(\lambda;T_{s}=5800 \textrm{ K}\right)$. The sites are composed of $N_{LHC}$ discrete LHCII trimers, each containing coupled \chlb and \chla domains (see Fig. \ref{Fig3} \textbf{a} and \textbf{b.}). We then subject this optimised antenna to spectral fluxes from lower mass stars and then attempt to recover photosynthetic performance by, as before, adding additional LHCII trimers. Finally, we apply the same approach to a modified antenna model which attempts to compensate for entropic barriers with an energetic 'funnel' structure. This is achieved by introducing a progressive blue-shift in $\lambda_{m}^{p}$ as more and more LHCII are added to the system.   

\section{Results}

\subsection{Orbital distances and incident spectral fluxes}

\begin{table}[!h]
\begin{center}
\caption{\label{tab:2} The maximum, $a_{sp}^{max}\left(T_{p}=273 \textrm{ K}\right)$, minimum, $a_{sp}^{min}\left(T_{p}=373 \textrm{ K}\right)$, and midpoint, $a_{sp}^{mid}$, habitable distances for stellar models defined by stellar temperature, $T_{s}$, and radius, $R_{s}$. Orbital distances are expressed in astronomical units, $1 \textrm{a.u.}\approx 1.50\times 10^{11}$ m, and stellar radii are expressed in terms of the Solar radius, $1 R_{\odot}\approx 6.96\times 10^{8}$ m}
\begin{tabular}{ |c|c|c|c|c| }
 \hline
 $T_{s}$ (K) & $R_{s} \left(R_{\odot}\right)$  & $a_{sp}^{max}$ (a.u.)& $a_{sp}^{min}$ (a.u.)& $a_{sp}^{mid}$ (a.u.) \\
 \hline
 \hline
$2300$& $0.117$& $0.019$& $0.010$& $0.015$\\
$2600$& $0.133$& $0.028$& $0.0.015$& $0.021$\\
$2800$& $0.254$& $0.062$& $0.033$& $0.048$\\
$3300$& $0.299$& $0.102$& $0.055$& $0.078$\\
$3800$& $0.613$& $0.276$& $0.148$& $0.212$\\
$4300$& $0.694$& $0.400$& $0.214$& $0.307$\\
$4800$& $0.775$& $0.557$& $0.299$& $0.428$\\
$5800$& $0.936$& $0.982$& $0.526$& $0.754$\\
\hline
\end{tabular}
\end{center} 
\end{table}

The habitable orbital distances, $a_{sp}^{min}\leq a_{sp}\leq a_{sp}^{max}$, derived using Eqn. (\ref{eq:2}) are listed in Table \ref{tab:2}. Note that Earth's orbit ($1 \textrm{a.u.} \sim 14.96\times 10^{10}$ m) lies slightly outside the predicted habitable zone for a $T_{s}=5800$ K star (our model for the Sun) due to neglect of any greenhouse effect. However, these distances are reasonable for qualitative comparison across a range of $T_{s}$. For the mid-point in this range we calculate $f_{p}\left(\lambda;T_{s}a_{sp}^{mid}\right)$ which are plotted in Fig. \ref{Fig1} \textbf{a}. In the range $4300 \textrm{ K} \leq T_{s} \leq 5800 \textrm{ K}$ the irradiance peak is in the PAR region, moving to the typical anoxygenic region ($750 \textrm{ nm}< \lambda < 1000 \textrm{ nm}$) for $3300 \textrm{ K} \leq T_{s} \leq 3800 \textrm{ K}$, and then red-shifting out of the window for photo-autotrophy on Earth for $T_{s}\leq 2800$ K. However, it should be noted that there is non-negligible irradiance in the PAR region for all $T_{s}$.   

\subsection{Thermodynamic antenna model: Increasing antenna size in limited PAR}

\begin{table}[!h]
\begin{center}
\caption{\label{tab:3} Kinetic and thermodynamic parameters for a thermodynamic antenna model optimized to $f_{p}\left(\lambda;T_{s}=5800\textrm{ K}\right)$. $N_{LHC}$ and $\tilde{\sigma}$ were treated as fit parameters while all others we derived from these and parameters in Table \ref{tab:1}, including $\nu_{e}$ and $\phi_{e}$ which are the output variables of the model. Energies are quoted in units of $k_{B}T$ assuming $T=300$ K.}
\begin{tabular}{ |c|c|c| }
 \hline
 Parameter & Denotes & Value \\
 \hline
 \hline
 $N_{LHC}$ & Number of LHC subunits & $5$\\
 $\tilde{\sigma}$& LHC absorption cross-section& $1.9\times10^{-18}$ m$^{2}$\\
 \hline
 $\Delta H_{1,2}$ & Enthalpy change, $N_{1}\rightarrow N_{2}$ & $-2.7k_{B}T$\\
 $T\Delta S_{i,j}$ & Entropy change, $N_{1}\rightarrow N_{2}$ & $+2.9k_{B}T$\\
 $\Delta F_{1,2}$ & Free energy change $N_{1}\rightarrow N_{2}$ & $-3.2k_{B}T$\\
 $\left(k_{1,2}^{-1},k_{2,1}^{-1}\right)$ & $N_{1}\leftrightharpoons N_{2}$ transfer times & $\left(1 \textrm{ ps}, 21 \textrm{ ps}\right)$\\
 \hline
$\Delta H_{2,RC}$ & Enthalpy change, $N_{2}\rightarrow N_{RC}$ & $-0.5k_{B}T$\\
 $T\Delta S_{2,RC}$ & Entropy change, $N_{2}\rightarrow N_{1}$ & $-4.9k_{B}T$\\
 $\Delta F_{2,RC}$ & Free energy change $N_{2}\rightarrow N_{RC}$ & $+4.4k_{B}T$\\
 $\left(k_{2,RC}^{-1},k_{RC,2}^{-1}\right)$ & $N_{2}\leftrightharpoons N_{RC}$ transfer times & $\left(712 \textrm{ ps}, 10 \textrm{ ps}\right)$\\
 \hline
 $\nu_{e}$ & Electron output rate & $97$ s$^{-1}$\\
 $\phi_{e}$ & Antenna quantum efficiency & $82$\\
 \hline
\end{tabular}
\end{center} 
\end{table}

For values $N_{LHC}=5$ and $\tilde{\sigma}=1.9\times 10^{-18}$ m$^{2}$ we obtain $\nu_e\sim 97$ s$^{-1}$ and $\phi_e \sim 82$ for $f_{p}\left(\lambda;T_{s}=5800\textrm{ K}\right)$ (the model Solar flux). For comparison, the integrated cross-section of \chla is approximately $\sigma \sim 9\time 10^{-20}$ m$^{2}$ \citep{Noy_Chla_sigma}. If we multiply this by $42$ (the number of Chl molecules in an LHCII trimer) then $\sigma_{LHCII}\sim 3.8\times 10^{-18}$ m$^{2}$ which is very close to our fit value. $N_{LHC}=5$ is broadly consistent with the structure of the PSII antenna in plants \citep{DEKKER200512}. This agreement implies that the structure of our model, the assumption of the steady state approximation, and the various experimental parameters are indeed reasonable. These fit parameters plus derived thermodynamic/rate parameters are listed in Table \ref{tab:2}. 

$N_{1} \rightarrow N_{2}$ energy transfer is thermodynamically favourable, since $N_{1}$ is higher in energy and contains fewer pigment states than $N_{2}$. $N_{2}\rightarrow N_{RC}$, however, is associated with a large entropic penalty of $T\Delta S_{2,RC}\sim -5k_{B}T$ making it very \emph{unfavourable}. Energy transfer \emph{to} the RC, $k_{2,RC}$, is two order of magnitude slower than tansfer \emph{back} into the antenna. Despite this light-harvesting is still remarkably efficient ($\phi_{e} \sim 0.8$) as constant light absorption in the outer antenna drives energy towards the RC and $k_{2,RC}^{-1}\sim 700$ ps is still significantly faster than excitation dissipation ($k_{diss}^{-1}\sim 4$ ns). 

Fig. \ref{Fig1} \textbf{b} shows $\nu_{e}$ and $\phi_{e}$ for the same model for different $T_{s}$. $\phi_{e}$ is independent of $T_{s}$ since it is purely determined by the balance between energy transfer, dissipation and trapping and not the photon input rates, $\gamma_{1}$ and $\gamma_{2}$. Conversely, $\nu_{e}$ is strongly determined by $\gamma_{1}$ and $\gamma_{2}$, and therefore falls off rapidly with decreasing $T_{s}$ as the overlap between $f_{p}\left(\lambda\right)$ and $G_{i}\left(\lambda\right)$ decreases.

In Fig. \ref{Fig1} \textbf{c} $N_{LHC}$ is gradually increased while keeping all other parameters the same. Since $\nu_{e}\sim 100$ s$^{-1}$ is the \emph{maximal} electron output rate of plant PSII, values greater then this are unphysical. The reason for this is that the RC can only \emph{process} one excitation at a time (with an approximate turnover time of $\sim 10$ ms) and while this is happening the RC is said to be ``closed'' \citep{Baker_2004}. We can capture this behaviour by enforcing $N_{e}\leq 1$ with a sigmoidal saturating function,
\begin{equation}\label{eq:27}
    k_{trap}\rightarrow k_{trap}\left(1-\frac{1}{1+\textrm{exp}\left[-b\left(N_{e^{-}-c}\right)\right]}\right)
\end{equation}
where $b$ and $c$ are arbitrary paramters chosen to make the sigmoidal behavior sufficiently sharp. However, Eqn. (\ref{eq:17}) must now be solved via numerical optimization which can lead to stability problems. Moreover, the effect is simply to cause $\nu_{e}$ to plateau at $100$ electrons s$^{-1}$ for increasing $N_{LHC}$ and one can achieve the same effect by simply correcting these unrealistic data points after the fact. As we will see, for lower values of $T_{s}$ (which are of particular interest in this work) $\nu_{e}\sim 100$ s$^{-1}$ is never achieved and RC saturation is not relevant. 

For $T_{s}\geq3300$ K $\nu_{e}=100$ s$^{-1}$ can be recovered simply by adding more LHCs to the antenna. However, for $T_{s}\leq 2800$ K the electron output saturates at $\nu_{e}<30$ s$^{-1}$, due to the entropic barrier effect. This is explicitly shown in Fig. \ref{Fig1} \textbf{d} where $\phi_{e}$ decreases with increasing $N_{LHC}$ and $\Delta F_{2,RC}$ increases logarithmically. Since both $\Delta F_{2,RC}$ and $\phi_{e}$ are independent of $T_{s}$ the \emph{diminishing returns} of increasing $N_{LHC}$ also apply to higher mass stars. While an organism can recover $\nu_{e}\sim 100$ s$^{-1}$ by increasing antenna size, the overall efficiency of the antenna only decreases.   

\subsection{Lattice antenna model: Increasing the size of a ``flat'' antenna in limited PAR}

\begin{figure*}[ht]
\centering
\includegraphics[width=\linewidth]{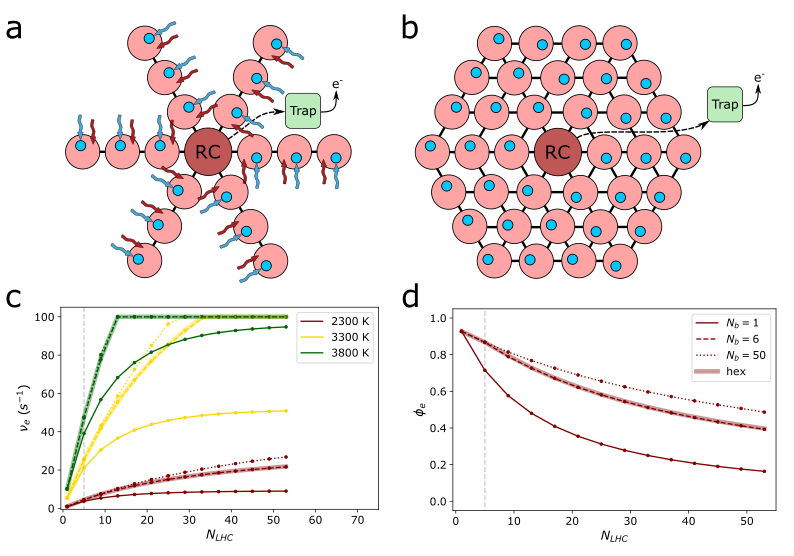}
\caption{\textbf{a.} Schematic of a \emph{branched} antenna, in this case composed of six branches each containing three LHC subunits. Each LHC has $N_1$ (blue circles) and $N_2$  (pink circles) domains, as in the previous thermodynamic model. $\nu_{e}$ can be increased by increasing the number of LHC subunits, $N_{LHC}$, and/or the number of branches, $N_{b}$. \textbf{b.} A hexagonal antenna in which LHCs are added sequentially in a clockwise manner from the central RC. \textbf{c.} $\nu_{e}$ as a function of $N_{LHC}$, $N_{b}$ and $T_{s}$. Solid lines indicate LHCs being added to a single branch ($N_{b}=1$), dashed lines indicate $N_{b}=6$ and dotted lines indicate $N_{b}=50$ (an upper limit where every new LHC is added to a new branch). The thick, faded curves are for a hexagonal antenna. The vertical dashed line indicates the $N_{LHC}=5$ antenna used to model plant PSII subject to $f_{p}\left(\lambda;T_{s}=5800 \textrm{ K}\right)$. The entropic barrier is somewhat lifted for a branched antenna of $N_{b}>1$. A hexagonal antenna performs as well as the $N_{b}=6$ branched antenna, indicating that the excitations take a direct path to the RC.
\textbf{d.} $\phi_{e}$ (which is independent of $\gamma_{1}$ and $\gamma_{1}$ and therefore independent of $f_{p}\left(\lambda;T_{s}\right)$). Increasing antenna size leads to an increase in the proportion of photons lost to dissipative processes in the antenna, reflecting the diminishing returns arising from the entropic limit.}
\label{Fig3}
\end{figure*}

The initial lattice model is a qualitative representation of plant PSII, with a set of identical antenna complexes arranged around a central RC (hence ``flat''). The sites are the \chlb and \chla domains of the LHCII trimers, with the \chlb domain connected only to the \chla domain within the same LHCII and inter-LHCII energy transfer involving only the \chla domains (see Fig. \ref{Fig3} \textbf{a} and \textbf{b}). We consider two antenna topologies: \emph{branched} and \emph{hexagonal}. In a branched antenna $N_{b}$ linear chains of LHCII radiate from the RC as shown in Fig. \ref{Fig3} \textbf{a} while the hexagonal antenna is a standard hex-lattice centred on the RC, shown in Fig. \ref{Fig3} \textbf{b}. For the branched antenna we consider $N_{b}=1, 6,$ and $50$. $N_{b}=1$ is directly comparable with our thermodynamic model, as there is only a single connection point between the antenna and RC, $N_{b}=6$ is chosen to be directly comparable with the hexagonal antenna, and $N_{b}=50$ is a maximally efficient (but unrealistic) model in which all LHCs (in the range considered) are directly connected to the RC. 

As before we initially optimize this model to $f_{p}\left(\lambda;T_{s}=5800\textrm{ K}\right)$. For consistency between the hexagonal and branched models we set $N_{LHC}=6$ and then tune $\tilde{\sigma}$, $K_{LHC,LHC}^{hop}$ and $K_{LHC,RC}^{hop}$, which are listed, along with $\nu_{e}$ and $\phi_{e}$ for each antenna topology, in Table \ref{tab:4}. All other parameters are the same as in the thermodynamic model. The inter-LHC hopping time was tuned to $\left(K_{LHC,LHC}^{hop}\right)^{-1}\sim 5$ ps which is a value used in previous models of energy transfer within the PSII supercomplex \citep{Broess_2006,Valkunas_CG_2009}. The LHC-RC hopping time, $\left(K_{LHC,RC}^{hop}\right)^{-1}\sim 18$ ps, is slightly slower which is reasonable given that in the supercomplex the LHCII trimers are connected to the RC by several minor LHC and core antenna complexes \citep{Broess_2006,Valkunas_CG_2009}. These values yield $\nu_{e}\sim 100$ s$^{-1}$ and $\phi_{e}\sim0.85$ for both a hexagonal antenna and a branched antenna with $N_{b}=6$. The $N_{b}=1$ antenna performs significantly worse ($\nu_{e}^{N_{b}=1}\sim 77$ s$^{-1}$ and $\phi_{e}^{N_{b}=1}\sim 0.65$), reflecting the fact that the excitation has a longer path to the RC and only a single entry point. 

\begin{table}[!h]
\begin{center}
\caption{\label{tab:4} Fit and kinetic parameters for (plant) PSII-like lattice models optimized to stellar flux $f_{p}\left(\lambda;T_{s}=5800\textrm{ K}\right)$. We consider $N_{LHC}=6$ in hexagonal and branched ($N_{b}=1$ and $6$) topologies.}
\begin{tabular}{ |c|c|c| }
 \hline
 Parameter & Denotes & Value \\
 \hline
 \hline
 $N_{LHC}$ & Number of LHC subunits & $6$\\
 $\tilde{\sigma}$ & LHC absorption cross-section & $1.6\times 10^{-18}$ m$^{2}$\\
 \hline
 $\left(K_{LHC,LHC}^{hop}\right)^{-1}$ & Inter-LHC hopping time & $5$ ps\\
 $\left(K_{LHC,RC}^{hop}\right)^{-1}$ & LHC-RC hopping time & $18$ ps\\ 
 \hline
 $\nu_{e}^{N_{b}=1}$ & Output rate ($N_{b}=1$ branched antenna) & $77$ s$^{-1}$\\
 $\phi_{e}^{hex}$& Efficiency ($N_{b}=1$ branched antenna) & $0.65$\\
 \hline
 $\nu_{e}^{N_{b}=6}$ & Output rate ($N_{b}=6$ antenna) & $100$ s$^{-1}$\\
 $\phi_{e}^{N_{b}=6}$ & Efficiency ($N_{b}=6$ antenna) & $0.85$\\
 \hline
 $\nu_{e}^{hex}$ & Output rate (hex antenna) & $100$ s$^{-1}$\\
 $\phi_{e}^{hex}$& Efficiency (hex antenna) & $0.85$\\
 \hline
\end{tabular}
\end{center} 
\end{table}

As before, $\nu_e$ decreases with $T_{s}$ and we then attempt to recover $\nu_{e}\sim 100$ s$^{-1}$ by increasing $N_{LHC}$. For clarity, we show only $T_{s}=2300$ K, $3300$ K, and $3800$ K, in Fig. \ref{Fig3} \textbf{c} as they exhibit qualitatively different behaviour. For $T_{s}\geq 3800$ K the hexagonal and $N_{b}=6$ antennae can fully recover $\nu_{e}\sim 100$ s$^{-1}$ by doubling antenna size ($N_{LHC}=10-15$), while the linear ($N_{b}=1$) antenna eventually recovers but requires a much larger increase. For $T_{s}=3300$ K the hexgonal and $N_{b}=6$ fully recover with a 6-fold increase in $N_{LHC}$ but the $N_{b}=1$ model plateaus at $\nu_{e}\sim 50$ s$^{-1}$. For $T_{s}=2300$ K the output plateaus at $\nu_{e} < 20$ s$^{-1}$, with the $N_{b}=1$ antenna being limited to $\nu_{e}<10$ s$^{-1}$ In Fig. \ref{Fig3} \textbf{d} we plot $\phi_{e}$ for various antenna models (independent of $T_{s}$) as a function of $N_{LHC}$. In all cases increasing $N_{LHC}$ decreases $\phi_{e}$, due to the increasing entropic barrier. The decreases is steepest for the $N_{b}=1$ antenna, dropping below $\phi_{e}=0.5$ at only $N_{LHC}\sim 10$. The decrease is much less severe for the hexagonal and $N_{b}=6$ antennae and, as with $\nu_{e}$, they are essentially identical. For comparison we also show an idealized $N_{b}=50$ model in which every LHC is connected only to the RC. This model is still subject to the increasing entropic barrier with increasing $N_{LHC}$, since transfer of energy back into the antenna becomes progressively easier. Interestingly, the $N_{b}=50$ model does not offer a significant improvement over the hexagonal and $N_{b}=6$ antennae.         

\subsection{Lattice antenna model: Improving antenna efficiency with an energetic 'funnel'}

\begin{figure*}[ht]
\centering
\includegraphics[width=\linewidth]{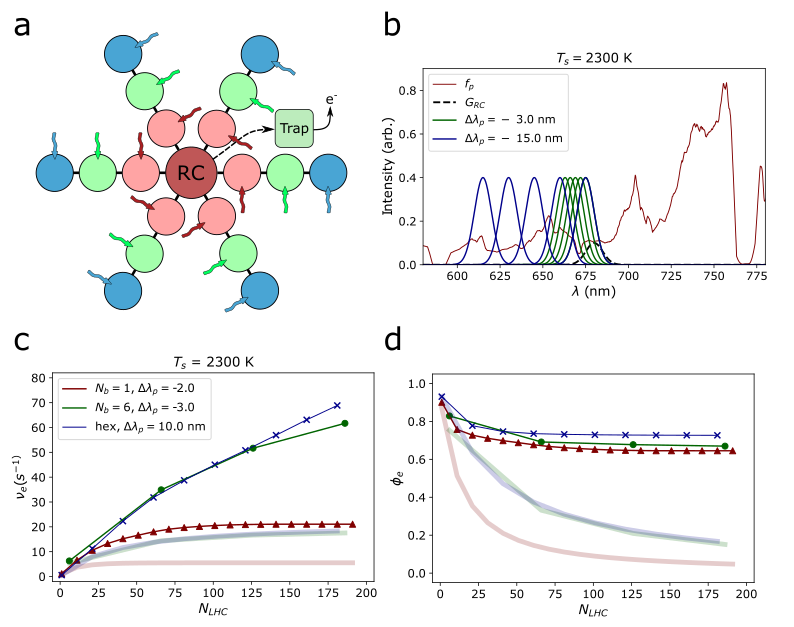}
\caption{\textbf{a.} Schematic of a 'funnel' antenna. While it still retains the hexagonal or branched structure of the the previous models, each LHC contains only one pigment type and there is a progressive blue shift in $\lambda^{p}$ as the LHCs get further from the RC (implied by the blue $\rightarrow$ red colour gradient). \textbf{b.} For illustrative purposes we show the absorption profiles (arbitrary scale) of a chain of LHCs with a small ($\Delta\lambda_{p}=3$ nm, green) and a large ($\Delta\lambda_{p}=15$ nm, blue) progressive blue-shift. For reference $f_{p}\left(\lambda;T_{s}=2300\textrm{ K}\right)$ is shown in red. The RC line-shape (arbitrary scale) is shown as a black, dashed line. \textbf{c.} $\nu_{e}$ as a function of $N_{LHC}$ for the optimal values of $\Delta\lambda_{p}$ for the $N_{b}=1$ (red), $N_{b}=6$ (green) and hexagonal (blue) antennae, for $f_{p}\left(\lambda;T_{s}=2300\textrm{ K}\right)$. The thicker, faded lines indicate $\Delta\lambda_{p}=0$ for the same models which are roughly comparable to the results in Fig. \ref{Fig3}. \textbf{d.} $\phi_{e}$ plotted for the same antenna topologies, parameters and $f_{p}\left(\lambda;T_{s}\right)$.}
\label{Fig4}
\end{figure*}

\begin{figure*}[ht]
\centering
\includegraphics[width=\linewidth]{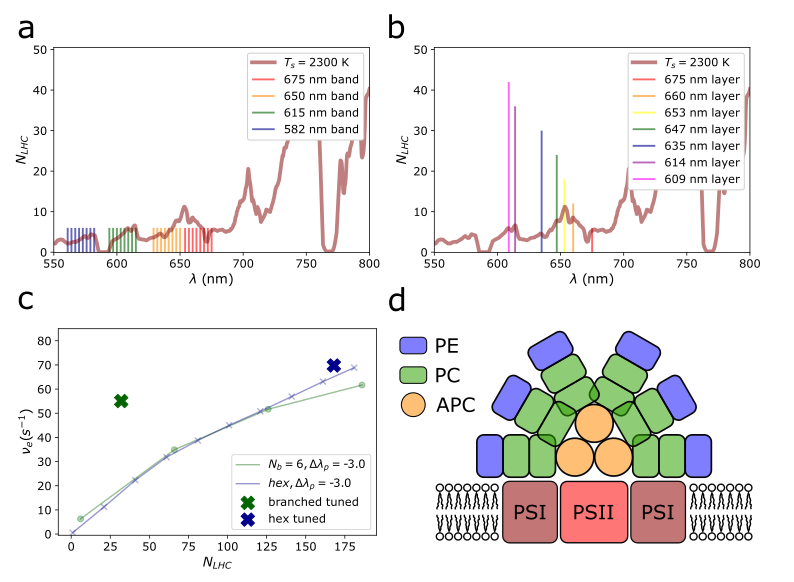}
\caption{\textbf{a.} The absorption maxima ($\lambda_{m}^{p}$, vertical lines) for a $N_{b}=6$ branched antenna specifically tuned to $f_{p}\left(\lambda;T_{s}=2300\textrm{ K}\right)$. The heights of the vertical lines indicate the relative proportion of LHCs of the different $\lambda_{m}^{p}$ which in this case are all equal. Bands of LHCs are tuned to local maxima in $f_{p}$, while small shifts between LHCs in the same band ensure that every transfer step is thermodynamically favourable. \textbf{b.} A hexagonal antenna tuned in the same manner. The outer concentric layers of the antenna contain more LHCs than inner layers. Coupled with the requirement for radial symmetry, this means that a limited number of $\lambda_{m}^{p}$ values can be considered. \textbf{c.} Optimal $\nu_{e}\left(N_{LHC}\right)$ for hexagonal (blue cross) and $N_{b}=6$ (green cross) antennae. The faded line are the curves from Fig. \ref{Fig4} \textbf{c} for comparison. The $N_{b}=6$ antenna can be substantially improved by spectral tuning, being more flexible and requiring less protein than the hexagonal antenna. \textbf{d.} A schematic diagram of the phycobilisome-PSII-PSI supercomplex from cyanobacteria (adapted from\cite{saer2017light}). PE, PC and APC denote the phycoerythrin, phycocyanin, and allo-phycocyanin-binding LHC subunits of the phycobilisome. Unlike the plant PSII antenna the phycobilisome sits on the membrane surface, docked to both the PSII and PSI RC complexes.}
\label{Fig5}
\end{figure*}

To compensate for the entropic barrier we modify the antenna so that each transfer step towards the RC is associated with a decrease in enthalpy (see Fig. \ref{Fig4} \textbf{a}). Retaining the same branched or hexagonal topologies, each LHC (which no longer resemble LHCII) is assumed to contain only one pigment type (as defined by $\lambda_{m}^{p}$). LHCs are added in concentric layers centred on the RC and a progressive blue-shift, $\Delta\lambda_{p}$, is introduced to the peak absorption wavelength, $\lambda_{m}^{p}$, of each layer. This produces a radially-symmetric energetic funnel centred on the RC (see Fig. \ref{Fig4} \textbf{b}). For brevity we consider only $T_{s}=2300$ K, this being the most light-limited case and it being simple to extrapolate the subsequent results to higher $T_{s}$. In the Supporting Material (Figs. S1-S3) we plot heatmaps of $\nu_{e}\left(N_{LHC},\Delta\lambda_{p}\right)$ and $\phi_{e}\left(N_{LHC},\Delta\lambda_{p}\right)$ for the hexagonal, $N_{b}=1$ and $N_{b}=6$ antennae, allowing us to identify the optimal $\Delta\lambda_{p}$ for each. These optimal $\nu_{e}\left(N_{LHC}\right)$ and $\phi_{e}\left(N_{LHC}\right)$ curves are shown in Figs. \ref{Fig4} \textbf{c} and \textbf{d.} respectively, alongside the curves for $\Delta\lambda_{p}=0$ which are roughly the same as those obtained for flat antennae in the previous subsection. Shown in Fig. \ref{Fig4} \textbf{c}, there is a substantial enhancement in $\nu_{e}$, particularly for the hexagonal and $N_{b}=6$ antennae. Interestingly, the hexagonal antenna requires $\Delta\lambda_{p}=-10$ nm while for the $N_{b}=6$ antenna $\Delta\lambda_{p}=-3$ nm, which implies that the hexagonal antenna has a slightly higher entropic barrier to overcome, due to the lateral energy transfer that is excluded in a branched antenna. Fig. \ref{Fig4} \textbf{d} shows that antenna efficiency does not decreases as sharply with increasing $N_{LHC}$ as with previous models and plateaus at $\phi_{e}=0.7-0.8$ for all models. This implies that the enthalpic funnel effect more-or-less compensates for the increasing entropic penalty of adding more LHCs.           

There is scope for additional fine-tuning of the funnel antennae to account for local peaks and troughs in $f_{p}\left(\lambda;T_{s}\right)$. Starting with $N_{b}=6$ we can achieve $\nu_{e}\sim 55$ s$^{-1}$ for $N_{LHC}=32$ by structuring the antenna into four \emph{bands} (see Fig. \ref{Fig5} \textbf{a}). Smaller progressive absorption shifts within each band ensure that every transfer step towards the RC is thermodynamically favourable. This represents a substantial improvement over the previous model. We are more limited in tuning the hexagonal antenna since $N_{LHC}$ increases non-linearly as we add more and more concentric \emph{layers} of LHCs. Moreover, the funnel requirement means that the antenna must be radially symmetric, meaning that all LHCs within the same layer must have the same $\lambda_{m}^{p}$. This means that adding more and more bands to the hexagonal antenna gets increasingly expensive in terms of LHC subunits. In Fig. \ref{Fig5} \textbf{c} we see that the tuned hexagonal antenna offers little improvement over the untuned one. These results imply that the branched antenna is more adaptable to the local spectral irradiance than the hexagonal antenna and, critically, can achieve similar $\nu_{e}$ with far fewer LHC subunits.    

\section{Discussion}

The results of our generalized antenna models suggest that the oxygenic photosynthesis is likely feasible on exo-planets orbiting low mass M-dwarf stars, though there are some restrictions. A light-harvesting system similar to PSII of vascular plants (but with a larger antenna) would be severely entropy-limited even before we factor in local limitations like water scarcity, atmospheric attenuation of light, etc. However, this reflects the fact that plants are evolved to deal with highly variable irradiance \citep{vialet2017importance,Li2021_Light_levels} rather than constant low light. One of the major risks to plants is actually excessive illumination which can cause oxidative damage to PSII, a metabolically costly phenomenon known as \emph{photoinhibition}, and one that should be familiar to anyone who has suddenly moved a houseplant into full sun. To mitigate photoinhibition \citep{aro_photoinhibition_1993}, plants possess regulatory mechanisms that suppress photon absorption and down-regulate the antenna \citep{ruban_nonphotochemical_2016,ruban2021mechanism}. Therefore it is unsurprising that such a system does not perform well in severely PAR-depleted light. However, by introducing an energetic gradient to the antenna (the energy funnel) we were able to significantly improve antenna efficiency and electron output. Of the branched and hexagonal antenna topologies the former performs better, able to match the performance of the latter but with fewer LHC subunits and a shallower energetic gradient. Fewer LHC subunits is perhaps the most important factor as it is directly related to the metabolic ``cost'' of the antenna. 

Of course, there are countless other modifications that could be made to the antenna. For example, one could increase pigment density which would increase the overall cross-section of each LHC and enhance inter-pigment energy transfer. The pigment concentration in LHCII is already very high, approximately $1$ mol dm$^{-3}$, and the protein scaffold enforces highly specific inter-pigment distances and relative orientations to avoid dissipative processes such as \emph{concentration quenching} \citep{liu2004crystal}. However, one could imagine an adaptation similar to the \emph{chlorosome} antenna of green sulphur bacteria \citep{oostergetel2010chlorosome}. This structure is a solid aggregate of $\sim 300,000$ BChls encapsulated in a membrane sheath, effectively the densest pigment packing possible, and it facilitates (anoxygenic) photosynthesis in extremely low illumination. It should be noted however that such as structure is is a significant metabolic investment and one that is not seen in any large multi-cellular organism. Alternatively, one could replace \chla and \textit{b} with pigments with larger intrinsic cross-sections and/or absorption widths, which would increase the rate of photon absorption, $\gamma_{i}$. Since intermolecular energy transfer depends on the oscillator strengths and spectral overlap of the pigments involved, this may also enhance the intrinsic transfer rates, $\left(K_{i,j}\right)^{-1}$. Still, the branched, funnel antenna strategy is a plausible one because it explicitly addresses the fundamental thermodynamic limits inherent in light-harvesting. More importantly, such an antenna actually exists on in Earth in the form of the \emph{phycobilisome} of oxygenic cyanobacteria \citep{saer2017light}. 

A sketch of the phycobilisome-PSII-PSI supercomplex is shown in Fig. \ref{Fig5} \textbf{d}. Unlike LHCII, the phycobilisome sits outside the plane of the membrane connected to PSII (and PSI) by \emph{Allophycocyanin} (APC) proteins which absorb at $\lambda_{APC}^{p}\sim 640-660$ nm. Radiating from the APC core are the antenna branches (often termed ``rods'') composed of \emph{phycocyanin} (PC, $\lambda_{PC}^{p}\sim620$ nm) and \emph{phycoerythrin} (PE, $\lambda_{PE}^{p}\sim570$ nm) proteins \citep{sil2022excitation}. There are some differences between our model and the phycobilisome. The pigments (various phycobilins) have much wider absorption profiles than the \chla and \textbf{b} $Q_{y}$-bands considered here and there is some evidence that energy transfer between the PE and PC sub-units at end of the rods is ultra-fast ($\sim 100$ fs) and possibly non-classical \citep{sil2022excitation}. However, we would (cautiously) suggest that the PE $\rightarrow$ PC $\rightarrow$ APC energetic structure was selected for by the entropic considerations we discussed. 

Like our model, the phycobilisome is modular and can adapt to changes in irradiance. \citet{Kolodny_2022_Phyco_blue} recently showed that cyanobacteria in deep water, where light is very blue-shifted, will lenghten their phycobilisome rods by adding more terminal PE subunits and will increase overall antenna efficiency, $\phi_{F}$, by tuning pigment content (and possibly density). The most important result, however, is the recent work of \citet{battistuzzi2023oxygenic} who showed that cyanobacteria could comfortably adapt to the the light typical of M-dwarf stars. Moreover, they used a sophisticated \emph{starlight simulator} setup which models atmospheric attenuation of light, meaning their cyanobacteria were likely subject to ever scanter illumination than we consider here. Interestingly, they also considered pure ``far red'' ($700-800$ nm) illumination and found that this retarded growth rate relative to the M-dwarf light, implying that, as in our model, the organism was adapting its antenna to intercept more photons in the depleted PAR region.It should be noted that they also considered on the red-adapted strains described in the introduction that fared equally well in M-dwarf and far red light.      

Strictly, both our modelling and the experimental work of \citet{Kolodny_2022_Phyco_blue}. concerns the \emph{acclimation} of an existing antenna to an unusual light-environment rather than the ground-up evolution of an antenna structure specifically optimized to it. Nevertheless, Kolodny's work demonstrates that oxygenic photosynthesis is, in principle, possible around very low mass stars and our findings indicate that the key consideration is overcoming a fundamental entropic barrier to light-harvesting. 

\section{Acknowledgements}

CDPD and SC would like to acknowledge support from the BBSRC London Interdisciplinary Bioscience PhD Consortium (LIDo). Additionally, CDPD would like to acknowledge support from BBSRC for funding work (grant BB/T000023/1) that developed the key ideas in this work. TJH is funded by a Royal Society Dorothy Hodgkin Fellowship. EG gratefully acknowledges support from the UK Science and Technology Facilities Council (STFC; project reference ST/W001047/1).

\section{Author Contribution Statement}

CDPD, TJH and in discussion with EG and CWM, devised the project. TJH and EG provided stellar spectral data. Light-harvesting model development and simulations were carried out by CDPD and SC. TJH, EG and CWM provided additional interpretation of results. All authors were involved in Manuscript drafting.

\section{Authors disclosure statement}

There are no conflicts of interest to disclose.

\bibliographystyle{plainnat}
\bibliography{references.bib}

\end{document}